# Common Source of Light Emission and Nonlocal Molecular Manipulation on the Si(111)-7×7 Surface

*Rebecca M. Purkiss[1], Henry G. Etheridge[1], Peter A. Sloan[1,2]\*, and Kristina R. Rusimova[1,2,3]\**

[1]Department of Physics, University of Bath, Bath, BA2 7AY, UK
[2]Centre for Nanoscience and Nanotechnology, University of Bath, Bath BA2 7AY, UK
[3]Centre for Photonics and Photonic Materials, University of Bath, Bath BA2 7AY, UK
\*E-mail: p.sloan@bath.ac.uk; k.r.rusimova@bath.ac.uk

Keywords: scanning probe microscopy, luminescence, Si(111)-7x7, desorption, nonlocal, hot electrons

The tip of a scanning tunnelling microscope can inject hot electrons into a surface with atomic precision. Their subsequent dynamics and eventual decay can result in atomic manipulation of an adsorbed molecule, or in light emission from the surface. Here, we combine the results of these two near identical experimental techniques for the system of toluene molecules chemisorbed on the Si(111)-7×7 surface at room temperature. The radial dependence of molecular desorption away from the tip injection site conforms to a two-step ballistic-diffusive transport of the injected hot electrons across the surface, with a threshold bias voltage of +2.0 V. We find the same threshold voltage of +2.0 V for light emission from the bare Si(111)-7×7 surface. Comparing these results with previous published spectra we propose that both the manipulation and the light emission follow the same hot electron dynamics, only differing in the outcome of the final relaxation step which may result in either molecular manipulation, or photon emission.

**Introduction**

The scanning tunnelling microscope (STM) in combination with an optical detection system can measure the light emission induced by charge tunnelling from the STM tip - an experiment called scanning tunnelling luminescence (STL).[1–3] In recent years, STL has been used to detect the light emission from single molecules,[4–10] the radiative decay of nano-cavity generated surface plasmons,[1,11,12] and even the production of pure single photons from molecular monolayers.[13] By employing the key capabilities of the STM for imaging with submolecular resolution and selectively exciting surface electronic states, STL has rendered itself the only far-field optical technique capable of circumventing the diffraction limit. By studying the interaction of light with individual molecules it has uncovered properties fundamentally different to those of molecular ensembles.[5,8–10,13] The three main mechanisms via which light is generated in the STL tunnel junction are (i) through the relaxation of a localised surface plasmon (LSP) in the region underneath the tip, (ii) by recombination of an electron-hole exciton,[14,15] or (iii) by electronic transitions.[6,7,16–19] The LSP model applies to metal surfaces, typically excited with a plasmonic (e.g. silver) STM tip, where the plasmon decays radiatively.[1,4,11,12] On the other hand, the exciton model commonly applies to decoupled single-molecule systems, where a molecular exciton can be excited either indirectly by the coherent coupling of a nanocavity plasmon, [5,8,9] or directly by the tunnelling electrons from the STM tip.[10,13] However, there are conflicting reports over the STL emission mechanism for semiconductor surfaces. In direct bandgap semiconductors, luminescence is caused by the recombination of tunnelling electrons with holes in the conduction band of the material.[18,19] In indirect bandgap semiconductors, where such recombination is forbidden, light emission is discussed in the context of a direct dipole transition of an electron from a tip state into a surface state, [16,17] or a radiative relaxation of a surface plasmon,[12,20] but the experimental evidence is contradictory.

Here we report both STL measurements and nonlocal molecular manipulation measurements on the Si(111)-7×7 semiconducting surface. Both experiments follow exactly the same experimental procedure, except that for STL we measure the light output of the hot-electron injection into the bare silicon surface, and for the atomic-manipulation we measure the reaction of adsorbed molecules on the surface.  In the latter case, toluene molecules adsorbed on the surface up to 20 nm distant from the STM tip are induced to leave the surface by an injection of electrons from the tip into the surface. We show that the voltage threshold of +2.0 V for such STM electron induced nonlocal atomic manipulation matches the voltage threshold for STL electron induced light emission.   Building on our prior nonlocal manipulation work[21–23] and other group's STL work,[20,24] we suggest a common mechanism for these electron induced events. Namely, a fast inelastic relaxation of the injected charge to the bottom of a high-lying electronic surface-resonance, followed by elastic diffusive transport away from the injection site and eventual inelastic decay to a lower lying surface state with  possible outcomes of molecule desorption, or of light emission.

**Results:**

*STM Induced Light Emission*

**Figure** 1(a) shows a typical STM image of the Si(111)-7×7 surface (+1 V, 100 pA, 25 × 25 nm). The repeating crystal structure is evident, as are several surface defects. The unit cell, outlined in Figure 1(a), contains six silicon adatom sites that correspond to the bright spots in the STM images. Due to a stacking fault of the surface, the unit cell has two sides: the faulted (F) and the unfaulted (U), each with two distinct adatom sites, the corner (C) and the middle (M). Hence we address, for example, the faulted corner silicon adatom site as FC. To measure the light emission we photographed the STM and sample system while the STM was in tunnelling position. Photographs of the tunnel junction were taken using an Andor Luca R camera, an electron multiplying CCD camera with near single photon sensitivity and a wavelength range of 400 nm to 900 nm (3.10 eV to 1.38 eV).  Figure 1(b) shows a photograph of the STM tip and sample with background illumination. This photograph

allows us to identify a small region of the camera image that contains the tunnel junction between tip and sample. With the background illumination switched off and light-tight covering applied to the UHV system we were able to record the light emission from the tunnel junction. Figure 1(c) shows the same camera region as the marked square of Figure 1(b), but with active STM tunnelling parameters of +3 V, 1 nA and no background illumination. A single pixel at the location associated with the tunnel junction has a much elevated photon count. We label this pixel as the `tunnel junction pixel' and describe these measurements as STL-photographs. We repeat the measurement directly after, but with the STM tip withdrawn by 10 nm away from the surface, thus ensuring no tunnelling current. These images we label as dark-photographs and they give a combined camera-bias and dark-count measurement. Here we present the raw camera counts, and in the final discussion we convert these counts to photons per injected electron.

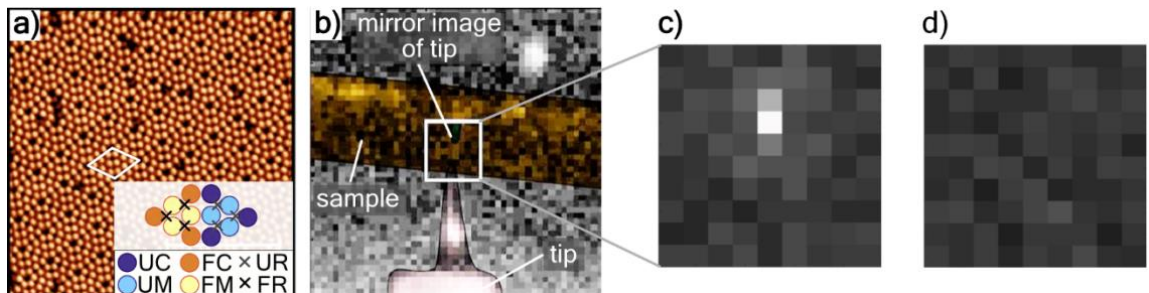

**Figure 1.** Detection of scanning tunnelling luminescence from the Si(111)-7×7 surface. (a) STM image of the Si(111)-7×7 surface (+1 V, 100 pA, 25 × 25 nm) with a unit cell outlined in white. Inset: Schematic of the outlined unit cell, showing all adatom and restatom sites as follows: unfaulted corner (UC); unfaulted middle (UM); faulted corner (FC); faulted middle (UM); unfaulted restatom (FR); faulted restatom (FR). (b) Camera image, with background lighting on, of the tunnel junction with silicon sample shaded orange, and the tip purple. The reflection of the tip can just be seen in the sample (best viewed online). The boxed region of interest contains the tip-apex/sample tunnel junction. (c) An STL-photograph (injection at +3 V, 1 nA, 40 s) of the boxed region of (b) with background light off, clearly indicating the enhanced light signal associated with the tunnel junction. (d) A dark-photograph (injection at +3 V, 0 nA, 40 s) of the boxed region of (b) also with background light off.

For each set of tunnelling parameters we typically record 11 pairs of STL and dark photographs. **Figure** 2(a) shows the number of camera-counts of the tunnel junction pixel over the course of 11 pairs of STL/dark photographs. There is an obvious elevation of the number of camera-counts when the tip is in tunnelling contact, but also a reasonable spread of camera-counts for both photographs: STL 1435 ± 98 and dark 642 ± 48 (error is one standard deviation of the mean, N=11). The camera-counts of a pixel some distance remote from the emission-pixel, Figure 2(b), shows little difference between STL (606 ± 46, N=11) and dark (660 ± 73, N=11) camera-counts. The fluctuation in the tunnel junction STL signal may be due to the high tunnelling current used in this experiment of 1 nA. At this relatively high tunnelling current the tip-apex can become unstable and change state during the tunnelling experiment leading to the degree of variation measured in the camera signal. It may also reflect the typically Poisson noise for our low photon count rate (see final discussion section).

Figure 2(c) presents the main STL result of this work, the voltage dependence of the light emission for electron injections from +1.5 V to +3 V. Each data point represents the average taken over typically 11 sets of STL/dark photographs. What is striking is the sharp voltage threshold at +2 V STM bias voltage. Below +2 V we find no light emission; above this threshold we find a near monotonically increasing light emission signal.

Previous STL reports on the Si(111)-7×7 surface have invoked both LSP and direct dipole transition mechanisms. In the work of Downes and Welland they reported a STL signal that lacked atomic-resolution and instead indicated a light emission region of ~2 nm surrounding their STM tip.[20] The light emission was modelled with a LSP in the tip/surface junction. Such a model would imply a photon-emission spectrum that was dependent on the geometric shape of the tunnel junction and its elemental identity.[25] On the other hand, the STL-spectroscopy work of Imada and co-workers,[24] showed atomic spacial resolution in the STL signal and no energy shift of the detected photon emission peaks with increasing tip-sample bias voltage. This lack of energy shift also rules out the previously proposed mechanism of a direct dipole transition between a tip state and a sample state,[17] which would produce a voltage dependence of the emission peak.

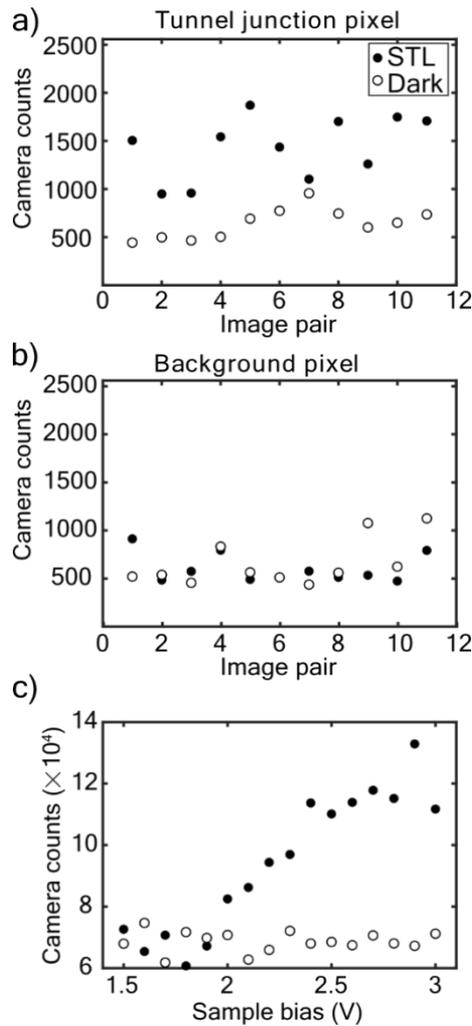

**Figure 2.** Light emission from the tunnelling junction pixel. (a) Camera-counts for the tunnel junction pixel over 11 pairs of STL photographs (Injection: +3 V, 1 nA, 40 s. Camera gain ×333) and dark photographs (Injection: +3 V, 0 nA, 40 s. Camera gain ×333). (b) Camera-counts taken from the same photographs as (a), but from a pixel some distance away from the tunnel junction pixel. (c) Voltage dependence of the STL and dark camera-counts (summed over 11 photographs) from the tunnel junction (Injections: 40 s. Camera gain ×833).

*STM Induced Nonlocal Atomic Manipulation*

There is no fundamental difference between an STL experiment and an atomic-manipulation experiment, except the measured outcome. In STL we measure the light emission generated by the hot-electrons injected from the tip of the STM. In atomic-manipulation we measure the response of molecules on the surface to the hot-electrons injected from the tip of the STM.

For each nonlocal atomic-manipulation experiment a large STM image was recorded with passive parameters (no molecular manipulation)[26] of a surface partly covered with chemisorbed toluene molecules which appear as dark-spots, see **Figure** 3(a). The tip was positioned at the centre of the image and a current injection performed with set voltage, current and time (here +2.8 V, 750 pA and 10 s). A second large STM image is then recorded, Figure 3(b), revealing a central region that has fewer molecular adsorbates (dark-spots) indicating a region of the surface that has undergone nonlocal molecular manipulation. That is, molecules some distance from the tunnel junction, i.e. injection site, have been induced to desorb from the surface.

We have previously reported on nonlocal manipulation of chlorobenzene and toluene from the Si(111)-7×7 surface.[21–23,27] See reference [28] for a general review of nonlocal manipulation. Here we use our ballistic-diffusive model, developed in references [22,23] to model hole-induced (negative bias voltage) nonlocal desorption, and apply it to electron-induced (positive bias voltage) nonlocal desorption. Our simple step-wise model proceeds as follows: (i) An electron tunnels from tip to surface and populates a 2D surface resonance; (ii) The wavefunction of this state evolves ballistically unperturbed for a certain time $\tau_i$ (of order 10 fs); (iii) The electron thermalises and relaxes to the bottom of the surface resonance band; (iv) it then undergoes a 2D random walk, i.e. 2D isotropic diffusion for ~ 200 fs; (v) the charge inelastically relaxes towards the Fermi level leading to the possible desorption of a molecule.

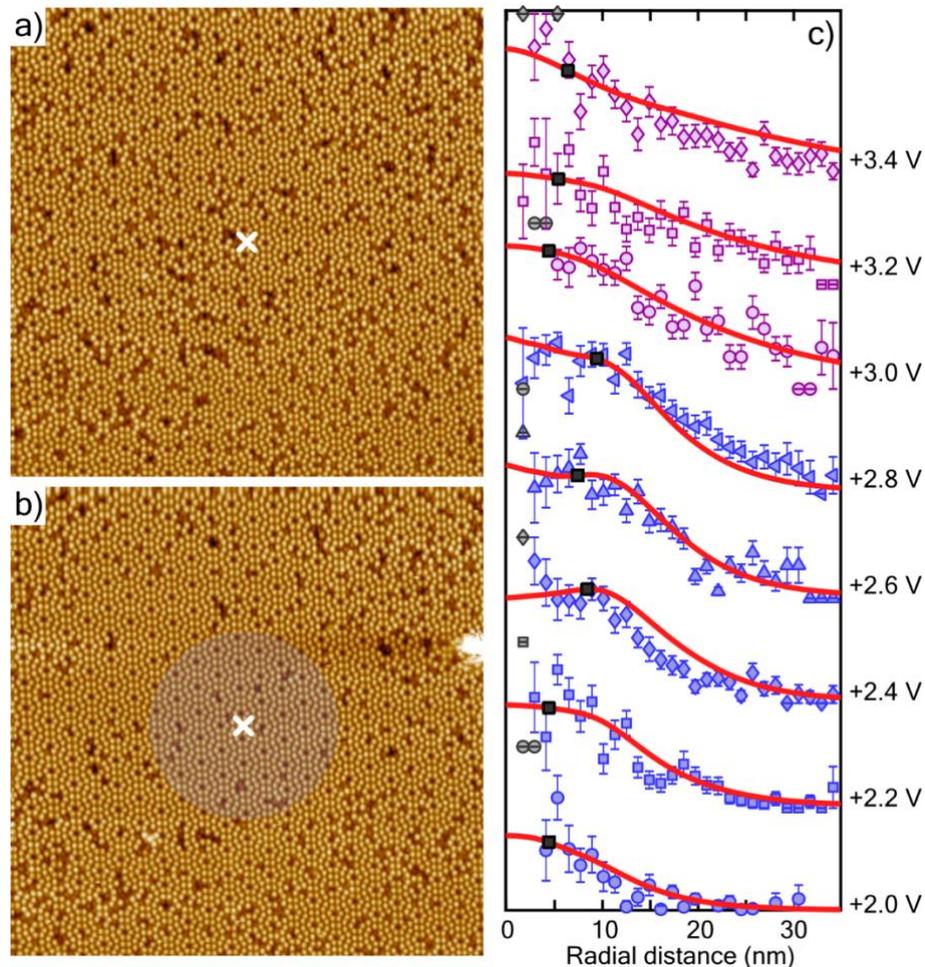

**Figure 3.** Nonlocal atomic manipulation with the STM. (a) Passive STM image (+1 V, 100 pA, 50 nm × 50 nm) taken before current injection. (b) STM image, with same parameters as (a), but taken after a current injection of +2.8 V, 750 pA and duration of 10 s located at the white ×. The shaded region indicates the measured hot electron diffusion length-scale of 10 nm for manipulation experiments at +2.8 V. (c) Injection voltage dependence of nonlocal manipulation of FM toluene molecules with a corner-hole injection site. Radial distribution curves have been vertically offset to aid clarity and match STS energy axis. Solid red lines show the inflation-diffusion model fitted to each dataset (light grey data points were omitted from the fits). Black markers indicate the range of the inflation region determined from the experimental data. Details of the fitting model are given in the main text and methods section. Data in blue/purple corresponds to electron transport by state $U_3$/$U_4$, respectively.

Figure 3(c) presents a series of curves showing the probability of manipulation as a function of distance from the injection site. From each curve we extract three key parameters: $β$ the probability per injected electron of inducing desorption; $\lambda$ the diffusive length-scale; and $R$ the region close to the tip where there appears to be a suppression of desorption. Here, in the `suppression' region, a purely diffusive model fails and the full ballistic-diffusive model is required. **Figure** 4 presents, in identical fashion to reference [23], these parameters as a function of the injection voltage as extracted from the experimental data curves of Figure 3(c).

Below +2.0 V we found no evidence of nonlocal manipulation. The same threshold we found for light emission. Between (+2.0 ± 0.1) V and (+2.9 ± 0.1) V we have a manipulation region of near constant diffusion length $\lambda$ = (14 ± 2) nm. Above this we enter into a second manipulation region with a diffusive length of $\lambda$ = (34 ± 4) nm. As in reference [23] we find a sharp reduction in the suppression region $R$ upon a change from one nonlocal regime to another. This is rationalised as the nonlocal transport changing from one surface electronic state to another. Hence, at the threshold voltage of +2.0 V, the electron wavepacket is generated at the bottom of a band and so has little or no group velocity and does not undergo any significant ballistic transport from the injection site before undergoing diffusive transport.

Scanning tunnelling spectroscopy (STS) measurements of a clean FM adatom, Figure 4(d), reveal the usual $U_3$ state at (+2.3 ± 0.5) eV with its onset responsible for the +2 V threshold for nonlocal manipulation. It also shows a higher-lying state at (+3.4 ± 0.3) eV, that we simply label for consistency as $U_4$. Hence, we label the first region as transport though the $U_3$ state as $T_{U_3}$, and the higher-lying region as $T_{U_4}$.

Using these STS parameters for the two regimes we can test whether our ballistic-diffusive model developed to model hole-injection, also describes electron-injection (see methods and reference [23] for detailed discussion and further justification). The fitted curves of Figure 3(b) show a good fit for radial distances greater than 10 nm away from the injection site, the length-scale dominated by diffusive transport. Below 10 nm we also find a good fit that describes the suppression of manipulation at these short ranges governed by the ballistic process. From a global fit to all the radial desorption curves within a region we extract an initial ballistic time for $T_{U_3}$ of $\tau_i = 5$ fs and the same for $T_{U_4}$ with $\tau_i = 5$ fs. We estimate a fitting error on the ballistic time of ~2 fs.

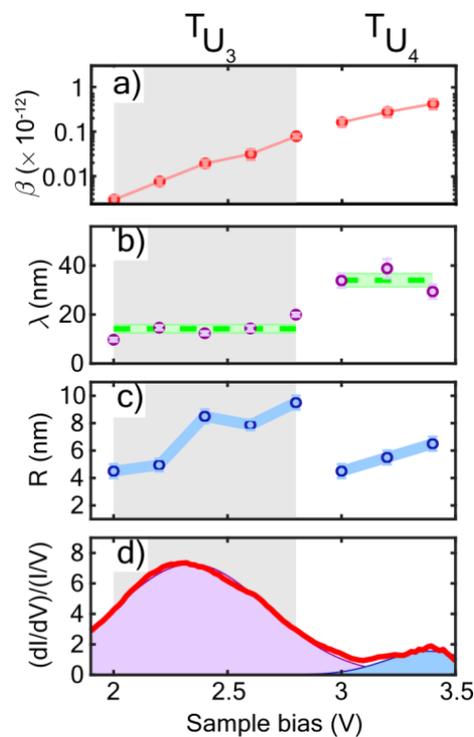

**Figure 4.** Injection bias dependence for electron injections into CH sites. Nonlocal manipulation of FM toluene molecules. (a) The probability $\beta$ of manipulation per injected hole. (b) Diffusion length-scale $\lambda$ with the average length-scale for each transport region indicated with green bar with uncertainty given by the width of the bar. (c) The range of the suppression region $R$. (d) Red curve, normalized STS spectrum taken on a clean FM site. Shaded areas show Gaussian fits to the two peaks with peak position and FWHM of $(2.3 \pm 0.5)$ eV and $(3.4 \pm 0.3)$ eV.

**Discussion**

The invariance of the nonlocal diffusive length-scale, Figure 4(b), to the injection voltage shows that once above a voltage threshold, all injected electrons are transported in the same fashion. That is, no matter their injection energy (voltage), the injected electrons are identical upon initiating the diffusive transport step of the nonlocal manipulation. Hence, after the initial ballistic step and before the diffusion, there must be an ultrafast relaxation of the charge to the bottom of its surface band.[29,30] As the STS of Figure 4(d) shows, for injections between +2 V and +3 V this surface band is the $U_3$ state, and above +3 V it is the $U_4$ state. The STL threshold found here of +2 V therefore suggests that if the same electron dynamics occur, then the light emission for injections above +2 V would be from the bottom of the $U_3$ state. We therefore propose that after the injected electron has undergone diffusion for 200 fs, [31] it decays out of the state and may emit a photon, or manipulate a molecule.

Therefore there are at least 3 decay channels: one leading to manipulation, one leading to light emission, and one that is inactive. Hence, within a transport regime, $T_{U_3}$ or $T_{U_4}$, the energy of the emitted photon should be voltage independent. Imada *et al.*[24] report just such a spectral invariance to the STL injection voltage. For electron injection at +2 V Imada *et al.* measure photon energy peaks at 1.4 eV and at 1.85 eV, and for electron injection at +3 V they measure photon energy peaks at 1.4 eV, 1.85 eV, and 2.4 eV, respectively. The invariance of the STL peaks to injection energy is analogous to our invariant nonlocal length scale $\lambda$ and agrees with having a common underlying physical process.

These results suggest that STL on the Si(111)-7 × 7 has the same initial ballistic-diffusive electron dynamics and that the electron has relaxed to the bottom of the high lying electronic band before decaying with the emission of a photon.

Taking the pre-photon emission energy as the energy threshold of our nonlocal manipulation we use the discrete energy losses observed in the spectra of Imada *et al.* to determine the final state of the injected electron upon its relaxation step. **Figure** 5 shows scanning tunnelling spectra, from 0 V to +3.5 V, of all four silicon adatom species of the surface. The $U_3$ state centred at (+2.3 ± 0.5) eV has a clear amplitude at each adatom site. The minimum energy required to manipulate a toluene molecule is +1.4 eV,[22,32] and the energy range of photons detectable by our camera is ≈ 1.4 to 3.1 eV. Thus, we would be insensitive to any transition between the bottom of the $U_3$ state and next state, $U_2$, observed at +1.3 V in the STS (ΔE = 0.7 eV). The next and lowest energy state is the dangling bond state $U_1$ which has slightly different values depending on the adatom species.[33] For middle adatoms the $U_1$ state lies at +0.3 eV. Hence a transition from the bottom of the $U_3$ band to the peak of the $U_1$ state gives an energy difference of ΔE = 1.7 eV. For corner adatoms, the $U_1$ state lies slightly higher at +0.5 eV giving an energy difference of ΔE = 1.5 eV.

Above +3 V injection bias we enter a new electron transport state $T_{U_4}$ which is predominantly located on the corner adatom sites, as STS shows in Figure 5. A transition to the $U_1$ state gives ΔE = 2.5 eV. For STL with +3 V injection Imada *et al.* find a spectral peak at 2.4 eV photon energy. We can therefore interpret the STL spectra of Imada *et al.* in terms of electron relaxation at particular adatom sites with initial energy at the bottom of the $U_3$ band, or as a cascade of relaxation steps from the U$_4$ band for electron injections above +3 V.

To quantitatively compare the two outcomes we recast both the manipulation and the light emission results into the probability of manipulation or photon emission per electron. The conversion from camera counts follows standard practice (see methods section). However, since the models used to describe the nonlocal manipulation[21–23] are based on individual electron-molecule interactions, we develop the following procedure to convert to the total manipulation probability. For simplicity we consider a purely diffusive model[22] and so the probability $P_m$ of manipulating a single molecule at radius $r$ is

$$P_m(r) = 1 - exp\left[-\frac{2n_i \beta \tau_D}{\pi \lambda^2} K_0\left(\frac{2r}{\lambda}\right)\right], \quad (1)$$

where $n_i$ is the number of injected electrons, $\beta$ is related to the probability of a single electron manipulating a single molecule, $\lambda$ the diffusive length-scale, $\tau_D$ the lifetime of the diffusive state, and $K_0$ a modified Bessel function of the second kind with argument $2r/\lambda$. Note in references [22] and [34] we used $\lambda = \sqrt{D\tau_D}$, here we use the correct form relating to 2D isotropic diffusion of $\lambda = \sqrt{4D\tau_D}$. Hence, Equation 1 is slightly modified from the version presented in [22,34]. For a single electron we can simply expand the exponential about zero to give $P_m(r) = \frac{2n_i \beta \tau_D}{\pi \lambda^2} K_0\left(\frac{2r}{\lambda}\right)$. Experimentally $P_m(r)$ is simply the ratio of the measured number of molecules in an annulus of radius $r$ before $N_0(r)$ and after $N(r)$ the electron injection, $P_m(r) = N(r)/N_0(r)$. Here, to compute the total number of manipulations that could happen in a single electron event we assume that each adatom site is manipulation `active' and hence $N_0(r) = \rho 2\pi r \Delta r$ for an annulus $\Delta r$, and with $\rho = 1.92 \times 10^{18}$ the number of adatoms per m². Therefore the total number of possible manipulation events $N$ for a single electron injection ($n_i = 1$) is $N = \int_0^\infty \rho \frac{2\beta\tau_D}{\pi\lambda^2} K_0\left(\frac{2r}{\lambda}\right) 2\pi r dr$. In the limit of $r \to \infty$ the integral $\int r K_0(2r/\lambda) dr \to \lambda^2/4$, hence $N = \rho\beta\tau_D$. We take $\tau_D$ = 200 fs[31] and so convert the $\beta$ values of Figure 4 to the total probability of manipulation per injected electron $N$ as displayed in Figure 5(c). This form of analysis should be identical to the original `scanning' manipulation work and indeed we find our results here match qualitatively and quantitatively with that earlier work.[27,35]

Figure 5(c) shows the voltage dependence of both the probability per electron of molecular manipulation and the probability per electron of photon emission. Both have a common +2 V threshold and then near exponential increase demonstrating near constant branching ratio between these two decay channels over the range from the threshold +2 V up to ~ +2.8 V. This similarity again suggests that, within a transport region, the state before molecular manipulation or light emission is the same common state no matter the injection voltage. This common state thus conforms to the proposal of a rapid relaxation of the injected charge to the bottom of a high lying state, before diffusive transport across the surface, and eventual decay out of that state with the possible outcomes of molecular manipulation or light emission.

As we demonstrated in an earlier work the actual molecular manipulation is in fact driven by excitation of the underlying silicon surface.[32] It is possible to manipulate the silicon adatoms of the surface in exactly the same fashion as reported here for the toluene adsorbates.[36,37] The main difference is whereas desorption of toluene is an irreversible process, the manipulated silicon adatoms are meta-stable at room temperature and drop back to their original sites before they can be imaged by an STM. We have also reported that the low lying state at ~ +0.5 V is reduced in intensity but still present at the location of the bonding adatom to a chlorobenzene adsorbate.[27] Toluene and chlorobenzene both bond and are manipulated in near identical fashion on the Si(111)-7×7 surface. Therefore, it is plausible that an inelastic transition from the surface state at +2 V to the dangling bond state at +0.5 V is responsible for both the light emission and the molecular manipulation. Figure 5(b) shows schematically our proposed relaxation process for electrons tunnelling from the STM tip into the sample. If the injected electrons have energy above the threshold for nonlocal manipulation, they first undergo ultrafast relaxation to the bottom of the $U_3$ state.[21,23] There they live for ~ 200 fs, during which they spread diffusively away from the STM injection site on the Si(111)-7×7 surface.[22] Finally, the electrons undergo relaxation to the dangling bond state ($U_1$). It is during this transition that photon emission, or molecular manipulation, may take place.

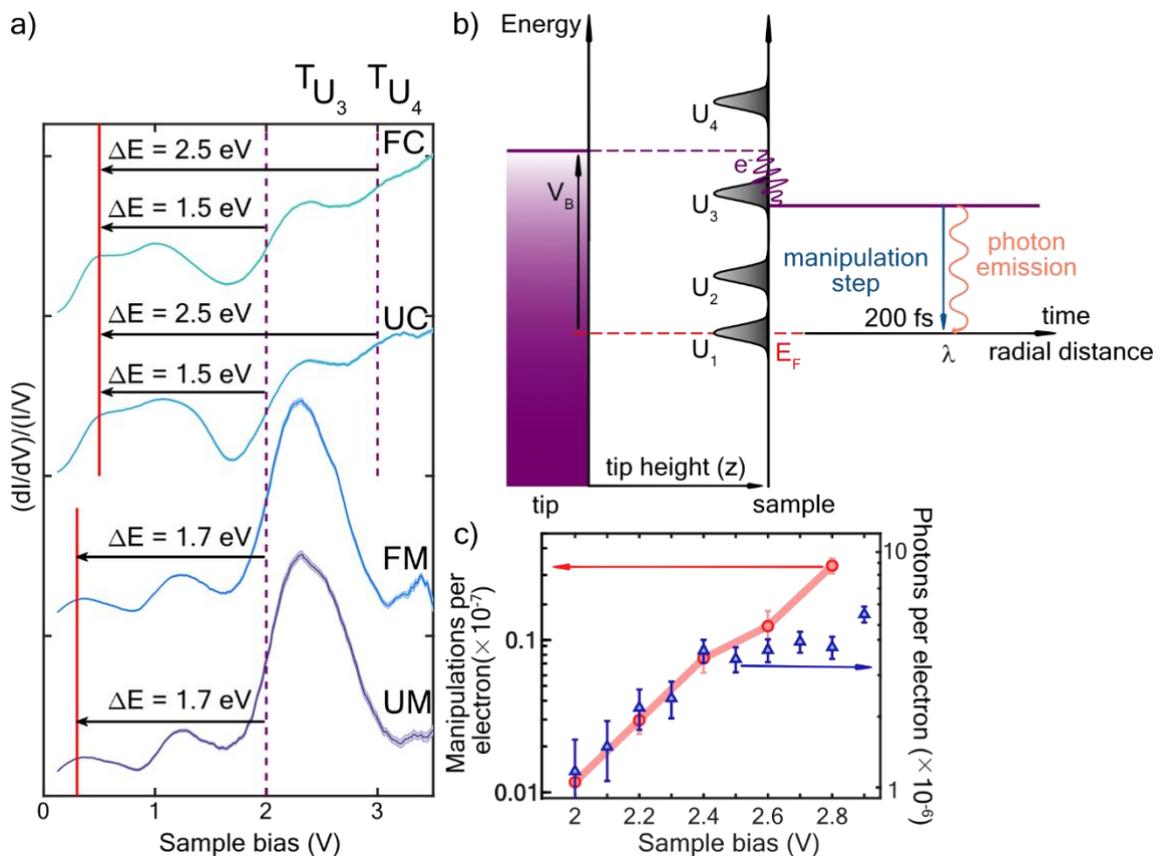

**Figure 5.** Proposed model for the energy relaxation of an electron injected from the STM tip into the sample. (a) Normalised density of states of the empty states. Each spectrum is the average of: 10, UM; 14, FM; 21, UC; 25, FC individual spectra, respectively. Spectra are offset vertically for clarity. Shaded area (barely visible) represents the standard error on each spectrum. Black arrows show the possible energy transitions for electrons on each site of the silicon surface. (b) Schematic showing the proposed energy relaxation mechanism for electrons tunnelling from the STM tip into the sample. An electron of energy higher than the nonlocal manipulation threshold first relaxes to the bottom of the $U_3$ state.[23] There it lives for ~ 200 fs, during which time it diffuses across the sample surface to a distance $\lambda$.[22] Next, the electron drops in energy into the $U_1$ state. During this final transition the electron may manipulate a molecule, or emit a photon. (c) Probability per electron of molecular manipulation (red circles) and probability per electron of photon emission (blue triangles) as a function of sample bias voltage.

**Conclusion**

The agreement between our luminescence threshold, our non-local manipulation thresholds, our STS and Imada e*t al.*'s[24] spectroscopy suggest a common origin for STL and nonlocal manipulation on the Si(111)-7×7 surface, namely the decay of an electron from the bottom of a high-lying band.

The combination of nonlocal manipulation with STL is a powerful experimental technique, which can be used to provide a deeper and more accurate understanding of the relaxation dynamics of hot electrons on a semiconducting surface. For the first time, we can access information about the final state of the relaxation dynamics of electrons injected from the tip of an STM. As a result, this technique opens up the possibility for tuning the density of surface states/resonances of a given semiconducting surface, which could lead to control over the lifetime of the excited state, and as a result over the size of the nonlocal effect. Thereby, we may be able to extend the lifetime (hence, the range) of hot electron transport, crucial to optimising both light harvesting devices and surface photochemical processes.

**Methods**

*UHV STM System*

Experiments were carried out on an Omicron STM1 under ultra-high vacuum with a base pressure less than 10$^{-10}$ mbar and operated at room temperature. Si(111) wafers (n-type, phosphorus doped, 0.001 to 0.002 Ω.cm) were prepared by repeated resistively heating to 1200° C followed by a slow cooling from 960° C to allow the surface to form 7×7 reconstruction.[38] Tungsten STM tips were produced by electrochemical etching of W-wire in 2M NaOH solution followed by a heat treatment in vacuum to remove any tungsten oxide. The STM was controlled by Nanonis digital control electronics.[39] Liquid toluene was purified by the freeze-pump-thaw technique with liquid nitrogen and checked for purity with a quadrupole mass spectrometer. To prepare a partially toluene covered surface (~3 molecules per unit cell) the Si(111)-7×7 surface was dosed through a computer-controlled leak valve.

*Scanning Tunnelling Spectroscopy*

Scanning tunnelling spectra were obtained by modulating the tunnelling voltage and recording d$I$/d$V$ directly with a lockin amplifier from 0 V to 3.5 V. The tunnel gap was varied linearly with voltage at a rate of 150 pm/V. Each spectrum contains 200 data points, collected at an integration time of 150 ms and lock-on modulation of 20 mV at 521 Hz. The STS curves in Figure 5 are the average of: 10, UM; 14, FM; 21, UC; 25, FC individual spectra, respectively.

*Luminescence measurement*

The Andor Luca R camera had 1004 (H) × 1002 (V) pixels with a pixel size 8 × 8 $\mu$m. To maximise the signal, but prevent pixel saturation, we used an exposure time $t$ of 40 s and a gain $G$ of 833. The camera was mounted at 80° to the surface normal and at a distance of ~

20 cm, giving a solid angle of ~ 0.024 sr. This solid angle is much lower, by at least an order of magnitude, than that typically used for STL experiments,[25] but given the single-photon sensitivity of the camera the solid angle was adequate. The angle is also appreciably away from the ideal 60° emission angle.[40] To convert from measured signal to emission probability per electrons $P_p$ we use the following conversion

$$P_p = \frac{(STL - Dark) \times W}{11 \times G \times t \times Q \times T \times S \times A \times n_e}, \qquad (2)$$

where: $STL$ is the raw camera counts summed over 11 single photographs; $Dark$ is the raw camera counts over 11 dark-photographs; $W$ = 50 is a tip-material factor to account for the low yield associated with tungsten; $G$ is the camera gain; $Q$ = 0.5 is the mean quantum efficiency of the camera; $T$ = 0.9 is the typical transmission of a glass UHV view port; $S$ is the ratio of camera solid angle to $4\pi$; $A$ = 0.3 is a correction for the angle of the camera away from the maximum light emission angle; $n_e = I/e$ is the number of electrons per second, which is simply related to the tunnelling current $I$ and the charge on an electron $e$. Although approximate this conversion should be within an order of magnitude or so of the true value. The critical conversion factor to allow quantitative comparison with other STL measurements is $W$. Here we take a conservative value.

The camera had a measured bias of 490 ± 30 counts and a dark count of 0.009 ± 0.001 per second per unit gain, giving a raw single dark-photograph camera-count of 490 + 0.009×$t$×$G$.

*Ballistic-diffusive model*

We employ the model from reference [23], which relates the magnitude of the suppression region $R$ to the surface band-structure. This allows us to extract the characteristic ballistic lifetime for each transport state. For each site, we have site-specific transport properties ($\lambda$, $\beta$, $R$) and spectroscopy measurements. This allows us to determine the site-specific transport region energy thresholds and band-widths. Using the s-band tight-binding model, the energy dispersion of the transport state is described by

$$E = E_0 + \frac{\Delta E}{2}(\cos(ka) - 1), \quad (3)$$

where $E_0$ is the energy threshold, $\Delta E$ is the state band-width and $a$ is the 7×7 unit cell size $a$ = 2.69 nm. From the site-specific plateau regions in $\lambda$, we determine energy onsets of the 2 transport states measured on FM's at (+2.0 ± 0.1) eV for the $U_3$ state and (+2.85 ± 0.1) eV for the $U_4$ state. The first threshold matches well with the position of the $U_3$ reported previously in [21,22,31]. The $U_3$ state has upwards dispersion.[41–43]

Previously, it has been measured that the $S_3$ state has a width of ~0.35 eV between the $\bar{M}$ point and half-way to the $\bar{\Gamma}$ point,[41,43] giving a full dispersion width of ~ 0.7 eV. The average FWHM of the $S_3$ state from our STS measurements[23] is 0.49 eV (0.54 eV for $U_3$; 1.08 eV for $U_4$). Based on the angle-resolved ultraviolet photoemission spectroscopy results, we calculate a conversion factor for the FWHM of the $S_3$ state,[41] giving a band-width of 0.7 eV for $S_3$ (and 0.8 eV for $U_3$; 1.5 eV for $U_4$). The state occupancy is then described with the standard equation for the tunnelling probability.[23,44]

**Acknowledgements**
PAS gratefully acknowledges support from the EPSRC grant EP/K00137X/1. RMP acknowledges funding and support from the Engineering and Physical Sciences Research Council (EPSRC) Centre for Doctoral Training in Condensed Matter Physics (CDT-CMP), Grant No. EP/L015544/1. The authors thank Dr Peter J. Mosley and Dr Kei Takashina for fruitful discussions. We thank Dr Peter J. Mosley for the use of the Andor Luca R camera. All data supporting this study are openly available from the University of Bath data archive at https://doi.org/10.15125/BATH-00622.